\documentclass[11pt]{article}
\usepackage{epsfig}
\usepackage{amsfonts}
\usepackage{amsmath}
\usepackage{bbm,bm}
\usepackage{cite}
\allowdisplaybreaks[1]

\newcommand{\nr}[1]{(\ref{#1})}

\newcommand{\fr}[2]{{\frac{#1}{#2}}}

\renewcommand{\vec}[1]{{\bf #1}}
\newcommand{\la}[1]{\label{#1}}
\newcommand{\be}{\begin{equation}}
\newcommand{\ee}{\end{equation}}
\newcommand{\ba}{\begin{eqnarray}}
\newcommand{\ea}{\end{eqnarray}}
\newcommand{\rmi}[1]{{\mbox{\scriptsize #1}}}

\newcommand{\ko}{\omega} %{k_0}

\newcommand{\eq}{eq.~}
\newcommand{\eqs}{eqs.~}
\newcommand{\se}{sec.~}
\newcommand{\ses}{secs.~}
\newcommand{\fig}{fig.~}
\newcommand{\figs}{figs.~}

\newcommand{\alphas}{\alpha_{\rm s}}
\newcommand{\Nf}{N_{\rm f}}
\newcommand{\Nc}{N_{\rm c}}
\newcommand{\Ns}{N_{\rm s}}
\newcommand{\Tc}{T_{\rm c}}
\newcommand{\rO}{r^{ }_0}
\newcommand{\tO}{t^{ }_0}

\newcommand{\Nt}{N^{ }_\tau}
\newcommand{\CF}{C_\rmii{F}}

\newcommand{\rmO}{{\mathcal{O}}}

\def\lsi{\raise0.3ex\hbox{$<$\kern-0.75em\raise-1.1ex\hbox{$\sim$}}}
\def\gsi{\raise0.3ex\hbox{$>$\kern-0.75em\raise-1.1ex\hbox{$\sim$}}}
\newcommand{\lsim}{\mathop{\lsi}}
\newcommand{\gsim}{\mathop{\gsi}}

\newcommand{\nF}{n_\rmii{F}}
\newcommand{\nB}{n_\rmii{B}}
\newcommand{\ff}{\rmi{\sl f\,}}
\newcommand{\f}{\mbox{\sl f\,}}
 \renewcommand{\nF}[1]{n_\rmii{F{#1}}}
 \renewcommand{\nB}[1]{n_\rmii{B{#1}}}
\newcommand{\rmii}[1]{{\mbox{\tiny\rm{#1}}}}

\newcommand{\Tint}[1]{{\hbox{$\sum$}\!\!\!\!\!\!\!\int\,}_{\!\!\!\!\raise-0.9ex\hbox{$\scriptstyle{#1}$}}}
\newcommand{\Tinti}[1]{{{\Sigma}\!\!\!\!\raise0.3ex\hbox{$\int$}_\rmii{${#1}$}}}

\newcommand{\bi}{\begin{itemize}}
\newcommand{\ei}{\end{itemize}}
\newcommand{\hide}[1]{ }

\makeatletter \@addtoreset{equation}{section} \makeatother
\renewcommand{\theequation}{\arabic{section}.\arabic{equation}}
\makeatletter
\renewcommand\section{\@startsection {section}{1}{\z@}%
                                   {-5.5ex \@plus -1ex \@minus -.2ex}% bfr-
                                   {2.3ex \@plus.2ex}%
                                   {\normalfont\large\bfseries}}
\renewcommand\subsection{\@startsection{subsection}{2}{\z@}%
                                     {-3.25ex\@plus -1ex \@minus -.2ex}%
                                     {1.5ex \@plus .2ex}%
                                     {\normalfont\normalsize\bfseries}}
\renewcommand\thesection {\@arabic\c@section}
\renewcommand\thesubsection   {\thesection.\@arabic\c@subsection}
\renewcommand{\@seccntformat}[1]{%
\csname the#1\endcsname.\hspace{1.0em}}
\makeatother
\begin{document}

\flushbottom

\begin{titlepage}

\begin{flushright}
BI-TP 2016/01 \\  
% HIP-2016-??/TH \\ % Notes M.L. \\ 
% http://arxiv.org/abs/1604.07544
July 2016
\end{flushright}
\begin{centering}

\vfill

{\Large{\bf
 Lattice constraints on the thermal photon rate
}} 

\vspace{0.8cm}

J.~Ghiglieri$^{\rm a}$, 
O.~Kaczmarek$^{\rm b}$, 
M.~Laine$^{\rm a}$, 
F.~Meyer$^{\rm b}$
% and 
% ... $^{\rm c}$

\vspace{0.8cm}

$^\rmi{a}$%
{\em
 AEC, ITP, University of Bern,  
 Sidlerstrasse 5, 3012 Bern, Switzerland\\
}

\vspace*{0.3cm}

$^\rmi{b}$%
{\em
       Faculty of Physics, University of Bielefeld, 
        33501 Bielefeld, Germany\\
}

% \vspace*{0.3cm}
% 
% $^\rmi{c}$%
% {\em
%        ... \\
% }

%        \email{okacz@physik.uni-bielefeld.de}, 
%        \email{laine@itp.unibe.ch},

\vspace*{0.8cm}

\mbox{\bf Abstract}
 
\end{centering}

\vspace*{0.3cm}
 
\noindent
We estimate the photon production rate from an SU(3) plasma at
temperatures of about $1.1\Tc$ and $1.3\Tc$. Lattice results for the
vector current correlator at spatial momenta $k\sim (2-6)T$ are
extrapolated to the continuum limit and analyzed with the help of a
polynomial interpolation for the corresponding spectral function,
which vanishes at zero frequency and matches to high-precision
perturbative results at large invariant masses. For small invariant
masses the interpolation is compared with the NLO weak-coupling
result, hydrodynamics, and a holographic model. At vanishing invariant
mass we extract the photon rate which for $k \gsim 3T$ is found to be
close to the NLO weak-coupling prediction. For $k \lsim 2T$
uncertainties remain large but the photon rate is likely to fall below
the NLO prediction, in accordance with the onset of a strongly
interacting behaviour characteristic of the hydrodynamic regime.

\vfill

%% %\noindent
%% %PACS numbers: 
%% %11.10.Wx, %        Finite temperature field theory
%% %11.15.Ha, %        Lattice gauge theory 
%% %12.38.Bx, %        Perturbative calculations in QCD
%% %12.38.Mh, %        Quark--gluon plasma
%% %12.39.Hg  %        Heavy quark effective theory 
%% %14.40.Nd, %        Bottom mesons
%% %\\
%% %Keywords: Thermal Field Theory,  NLO Computations, Lattice QCD
%% 

%% \vspace*{1cm}
  
\vfill

\end{titlepage}

%%%%%%%%%%%%%%%%%%%%%%%%%%%%% SECTION %%%%%%%%%%%%%%%%%%%%%%%%%%%%%%%%%%%%
%
\section{Introduction} 

The intensity and spectral properties of the photons that are
emitted from a thermal QCD plasma constitute excellent probes for 
the interactions that the plasma particles experience. Consequently, 
observing a thermal component in the photon yield of 
heavy ion collision experiments is among the main goals of 
the on-going program~\cite{rhic1,rhic2,alice}. 
Simultaneously, on the 
theory side, the thermal photon rate has served as a classic 
testing ground for developing increasingly  
advanced computational tools 
\cite{old1,old2,old3,htl2,htl3,photon1,photon2,ak}.

In order to test thermal QCD in a model-independent way, we would like
to compare first-principles computations with experimental
heavy-ion data. Apart from difficulties related to large
non-thermal backgrounds, 
this goal is faced with formidable challenges  
on the theory side.
On one hand, 
QCD continues to be strongly coupled 
in the temperature range reached in practice, 
so that a weak-coupling expansion may not suffice for
obtaining quantitatively accurate predictions
(unless a very high order is reached, cf.\ e.g.\ ref.~\cite{mDebye}). 
On the other hand, lattice
QCD is not directly applicable either, because simulations are
carried out in Euclidean spacetime, and analytic continuation
to Minkowskian signature represents a numerically ill-posed problem
(though the problem is again surmountable in principle~\cite{cuniberti}).

In the present paper, we suggest and test 
a pragmatic workaround to these
challenges, which could lead to 
a relatively reliable practical estimate of the photon production rate 
in the temperature range accessible to the current generation of heavy
ion collision experiments. The idea is to combine lattice and perturbative 
techniques, but only in regimes where they should be well under control. 
Concretely, this means that we make use of the weak-coupling expansion
in the regime of large ``photon masses'', ${M} \gsim \mbox{1~GeV}$, 
where the series shows reasonable convergence thanks to asymptotic freedom and
the high loop order that has been reached. 
This ``hard'' component permits for us
to reproduce the continuum-extrapolated lattice measurements at small
imaginary-time separations. In contrast, at large imaginary-time
separations the lattice data show clear deviations from the 
weak-coupling prediction. In order to account for these, we suggest
a general polynomial description 
of the spectral shape at ``soft'' photon masses. The parameters of 
the interpolation are determined through a least-squares fit to the
lattice data at large imaginary-time separations. Subsequently 
the fit result can be employed in order to extract spectral 
information concerning the soft domain. 

This paper is organized as follows. 
After discussing what is known theoretically about the vector
channel spectral function in various regimes in \se\ref{se:theory}, 
we introduce a general polynomial interpolation, designed to describe
the soft regime, in \se\ref{se:interp}.
The lattice analysis, incorporating a continuum extrapolation 
at three non-zero momenta and two temperatures, 
is described in \se\ref{se:latt}. Our fitting strategy and the 
corresponding results are presented in \se\ref{se:fit}, 
and we conclude in \se\ref{se:concl}. In an appendix 
the analysis is repeated for lattice data at zero momentum, 
pointing out that systematic uncertainties are much larger in this case. 

%%%%%%%%%%%%%%%%%%%%%%%%%%%%% SECTION %%%%%%%%%%%%%%%%%%%%%%%%%%%%%%%%%%%%
%
\section{Theoretical constraints on the vector channel spectral function} 
\la{se:theory}

%%%%%%%%%%%%%%%%%%%%%%%%%%%%% SECTION %%%%%%%%%%%%%%%%%%%%%%%%%%%%%%%%%%%%
%
\subsection{Basic definitions}
\la{ss:basic}

To leading order in the electromagnetic fine structure constant but
to all orders in the strong coupling,  the photon production rate 
per unit volume can be expressed as~\cite{text1,text2}
\ba
 \frac{{\rm d}\Gamma_\gamma(\vec{k})}{{\rm d}^3\vec{k}}
 & = & 
 \frac{1}{(2\pi)^3 2 k}
 \sum_{\lambda} 
 \epsilon_{\mu,\vec{k}}^{(\lambda)}
 \epsilon_{\nu,\vec{k}}^{(\lambda)*} 
 \int_{\mathcal{X}}  
  e^{i\mathcal{K}\cdot\mathcal{X}}
 \big\langle 
    J^{\mu}_\rmi{em} (0) 
    J^{\nu}_\rmi{em} (\mathcal{X}) 
 \big\rangle^{ }_{ }
 \\ & = & 
  \frac{1}{(2\pi)^3 2 k}
 \int_{\mathcal{X}}  
  e^{i\mathcal{K}\cdot\mathcal{X}}
 \big\langle 
    \sum_{i=1}^3 J^{i}_{\rmi{em}} (0) 
    J^{i}_\rmi{em} (\mathcal{X}) 
 - 
    J^{0}_{\rmi{em}} (0) 
    J^{0}_\rmi{em} (\mathcal{X}) 
 \big\rangle^{ }_{ }
 \;, \la{photon_rate}
\ea
where 
 $\mathcal{K} \equiv (k,\vec{k})$, $k \equiv |\vec{k}|$;
 $\mathcal{X} \equiv (t,\vec{x})$;
 $\mathcal{K}\cdot\mathcal{X} \equiv k t - \vec{k}\cdot\vec{x}$,   
 $ \epsilon_{\mu,\vec{k}}^{(\lambda)} $ 
denote polarization vectors, 
and $J^{\mu}_\rmi{em}$ is the electromagnetic current. 
In the second step we made use of a Ward identity, guaranteeing
that longitudinal polarizations do not contribute for $\mathcal{K}^2 = 0$. 

The electromagnetic current can in turn be expressed as 
$
 J^{\mu}_\rmi{em} = e 
  \sum_{\ff=1}^{\Nf} Q_{\ff}^{ } V^{\mu}_{\ff} 
$, 
where 
$
 V^{\mu}_{\ff} \equiv \bar{\psi}^{ }_{\ff} \gamma^{\mu} 
 \psi^{ }_{\ff}
$
is the vector current associated with the quark flavour {\sl f},
and $Q^{ }_{\ff}$ denotes the electric charge of flavour $\f$ in units of
the elementary charge $e$.
We consider the case of three degenerate flavours, $\Nf = 3$, 
so that $\sum_{\ff=1}^{\Nf} Q_{\ff}^{ } = 0$ and
$\sum_{\ff=1}^{\Nf} Q_{\ff}^{2} = 2/3$. 
Then the disconnected
quark contraction drops out. Relating furthermore the
Wightman correlator of \eq\nr{photon_rate} to a spectral function 
we can write 
\be
  \frac{{\rm d}\Gamma_\gamma(\vec{k})}{{\rm d}^3\vec{k}}
  \; = \;
  \frac{e^2 \sum_{\ff=1}^{\Nf} Q_{\ff}^2 }{(2\pi)^3 k}
 \, \nB{}(k) \,\rho_\rmii{V}^{ }(k,\vec{k})
 \;, \la{photon}
\ee
where $\nB{}$ is the Bose distribution. 
The vector channel spectral function has been defined as 
\be
 \rho_\rmii{V}^{ }(\omega,\vec{k}) 
 \; \equiv \;
 \int_{\mathcal{X}}  
  e^{i(\omega t - \vec{k}\cdot\vec{x})}
 \Big\langle 
    \fr12 \bigl[
     V^{i}_{ } (t,\vec{x}) \, , \,
     V^{i}_{ } (0) 
    \bigr]
 -
    \fr12 \bigl[
     V^{0}_{ } (t,\vec{x}) \, , \,
     V^{0}_{ } (0) 
    \bigr]
 \Big\rangle^{ }_\rmi{c}
 \;, \la{rhoV}
\ee
where $\langle...\rangle_\rmi{c}$ indicates that only the connected contraction
is included. The same spectral function also determines the dilepton
production rate as
\ba
 \frac{{\rm d} \Gamma_{\ell^-\ell^+}(\omega,\vec{k})}
   {{\rm d}\omega\, {\rm d}^3\vec{k} } & = &   
 \frac{ 2 e^4 \sum_{\ff} Q_{\ff}^2 \, \theta(M^2 - 4 m_\ell^2) } 
  {3 (2\pi)^5 M^2} 
 \biggl( 1 + \frac{2 m_\ell^2}{M^2}
 \biggr)
 \biggl(
 1 - \frac{4 m_\ell^2}{M^2} 
 \biggr)^\fr12 n^{ }_\rmii{B}(\omega) \, 
 \rho_\rmii{V}^{ }(\omega,\vec{k})  
 \;, \hspace*{5mm}
 \la{dilepton}
\ea
where the invariant mass of the dilepton pair has been defined as
\be
 M^2 \; \equiv  \; \omega^2 - k^2
 \;. \la{M}
\ee

%%%%%%%%%%%%%%%%%%%%%%%%%%%%% SECTION %%%%%%%%%%%%%%%%%%%%%%%%%%%%%%%%%%%%
%
\subsection{NLO weak-coupling expansion}
\la{ss:nlo}

In vacuum ($T=0$, where $T$ denotes the temperature), 
$ \rho_\rmii{V}^{ } $ is a function only of the photon 
invariant mass defined in \eq\nr{M}. 
The presence of a thermal plasma breaks Lorentz invariance, so that 
$ \rho_\rmii{V}^{ } $ is a function of two independent kinematic
variables, $\omega \pm k$. In particular, 
in the non-interacting limit~\cite{ga}, 
\be
 \rho_\rmii{V}^{ }(\omega,\vec{k}) \; = \; 
 \frac{\Nc T M^2 }{2\pi k}
 \biggl\{
 \ln\biggl[
  \frac{\cosh(\frac{\omega + k}{4T})}{\cosh(\frac{\omega - k}{4T})}
 \biggr]
   - \frac{\omega\,\theta(k-\omega)}{2T}
 \biggr\}
 \;, \la{free}
\ee
where $\Nc = 3$. This ``Born'' or 
``thermal Drell-Yan'' rate provides for a reasonable approximation
at large invariant masses, $M \gg \pi T$. However
for zero invariant mass the 
Born rate vanishes, and the leading-order (LO) result is proportional
to $\alphas T^2$.  

The determination of the correct LO result poses 
a formidable challenge~\cite{photon2}. 
However there is a logarithmically enhanced term that can 
be worked out analytically~\cite{htl2,htl3}, 
\be
 \rho_\rmii{V}^{ }(k,\vec{k}) \; = \;
  \frac{\alphas \Nc \CF T^2}{4} 
  \ln \biggl( \frac{1}{\alphas} \biggr)
  \Bigl[ 1 - 2 \nF{}(k) \Bigr] + \rmO(\alphas T^2)
  \;, \la{log}
\ee
where $\nF{}$ is a Fermi distribution and 
$\CF \equiv (\Nc^2 - 1)/(2\Nc)$. 
The non-logarithmic terms
are only known in numerical form~\cite{photon1,photon2}. Recently, 
these results have been extended to 
$
 \rmO(\alphas^{3/2} T^2)
$ 
both at vanishing~\cite{ak} and
non-vanishing photon masses ($|M| \lsim gT$, 
where $g \equiv \sqrt{4\pi\alphas}$)~\cite{gm}. 
In the following we make use of the results of ref.~\cite{gm}.

If the photon mass is large, $M\gg g^{1/2}T$, then 
there is a ``crossover'' to a different type of behaviour~\cite{lpm,gm}. 
For $M \sim \pi T$ the NLO corrections are 
suppressed by $\alphas$ and numerically small~\cite{master,dilepton}.
For $M \gg \pi T$, the spectral function goes over into a vacuum
result~\cite{ope} which is known to relative accuracy 
$\rmO(\alphas^4)$~\cite{kit_ns,kit_si} and can directly be taken 
over for a thermal analysis~\cite{cond4,dilepton}.
Such precisely determined 
results play an essential role in our investigation. 

%%%%%%%%%%%%%%%%%%%%%%%%%%%%% SECTION %%%%%%%%%%%%%%%%%%%%%%%%%%%%%%%%%%%%
%
\subsection{Hydrodynamic regime}
\la{ss:hydro}

A special kinematic corner in which it is possible
to make statements about $\rho^{ }_\rmii{V}$ beyond the weak-coupling
expansion is given by the so-called
hydrodynamic regime, parametrically $\omega,k \lsim \alphas^2 T$.
This is the regime in which the 
general theory of statistical
fluctuations~\cite{landau9} applies.  
Then the properties of $\rho^{ }_\rmii{V}$ can be parametrized
by a diffusion coefficient, denoted by $D$, and by a susceptibility,
denoted by $\chi^{ }_\rmi{q}$. The susceptibility determines the value of the
conserved charge correlator at zero momentum, 
$
 \chi^{ }_\rmi{q} \equiv \int_0^{\beta} \! {\rm d}\tau \int_{\vec{x}}
 \langle V^0(\tau,\vec{x}) V^0(0) \rangle
$,
whereas $D$ can be defined
through a Kubo formula as 
\be
 D_{ } \equiv 
 \fr1{3 \chi^{ }_\rmi{q}} 
 \lim_{\omega\to 0^+ } 
 \sum_{i=1}^{3}
 \frac{\rho^{ii}_{ }(\omega,\vec{0})}{\omega}
 \;. \la{Kubo_D}
\ee
The electrical conductivity is a weighted sum
over these quantities, 
\be
  \sigma =  e^2 
  \sum_{\ff=1}^{\Nf} Q_{\ff}^2 
  \chi^{ }_\rmi{q} 
  D^{ }_{ }
 \;, \la{sigma}
\ee
where the disconnected contribution has been omitted thanks to
$ \sum_{\ff} Q_{\ff}^{ } = 0 $.
% (if non-zero quark masses were kept, $\chi^{ }_\rmi{q}$ and $D$ could
% depend on the flavour index $\f$). 

In the hydrodynamic regime, the full 
$\rho_\rmii{V}^{ }$ can be expressed in terms
of $D$ and $\chi^{ }_\rmi{q}$. As already mentioned
the longitudinal components do not contribute at the on-shell
point, but they have a non-trivial diffusive structure elsewhere, 
leading to the prediction (cf.\ e.g.\ ref.~\cite{llog})
\be
 \frac{\rho_\rmii{V}^{ }(\omega,\vec{k})}{\omega} = 
 \biggl( \frac{\omega^2 - k^2} {\omega^2 + D^2 k^4}
  + 2 \biggr) \chi^{ }_\rmi{q} D
 \;. \la{hydro}
\ee
Consequently the photon production rate from \eq\nr{photon} becomes
\be
 \frac{{\rm d}\Gamma_\gamma(\vec{k})}{{\rm d}^3\vec{k}}
 \; \stackrel{k \lsim \alphas^2 T}{\approx} \; 
 \frac{2 T\sigma}{(2\pi)^3 k} 
 \;. \la{photon_res_1}
\ee
We also note that, for $\omega \ll D k^2$
and $k \ll 1/D$, \eq\nr{hydro} predicts that 
\be
   \lim_{\omega\to 0} 
   \frac{\rho_\rmii{V}^{ }(\omega,\vec{k})}{\omega} =  
 - \frac{\chi^{ }_\rmi{q}}{ D k^2}
 \;, \la{pred}
\ee
i.e.\ the slope  
should be negative at small enough frequencies.
The reason is that for very small $k$, $\rho^{00}_{ }$ resembles
a Dirac delta-function, which comes with a negative sign 
in $\rho^{ }_\rmii{V}$.\footnote{%
 The physical spectral function is positive 
 at and somewhat below the light 
 cone~\cite{gm}.  According to \eq\nr{pred}, 
 it should cross zero at some $\omega < k $ if 
 $k$ is small enough, $k \lsim \alphas^2 T$. Because of unknown 
 numerical prefactors, 
 it is unclear whether such $k$ are reached in our
 simulations. 
 }  

%%%%%%%%%%%%%%%%%%%%%%%%%%%%% SECTION %%%%%%%%%%%%%%%%%%%%%%%%%%%%%%%%%%%%
%
\subsection{AdS/CFT limit}
\la{ss:ads}

In the AdS/CFT framework $\rho_\rmii{V}^{ }$
has the same infrared structure as in \eq\nr{hydro},  
with the specific values 
$D = 1/(2\pi T)$ and $\chi^{ }_\rmi{q} = \Nc^2 T^2/8$~\cite{ads1,ads2}. 
The spectral function is close to the hydrodynamic
form for $k \lsim 0.5/D$, and becomes negative at the smallest
$\omega$ for $k\lsim 1.07/D$.
Below we make use of the results of ref.~\cite{ads2}, evaluated numerically
so that they make predictions beyond the hydrodynamic regime as well. 
Of course, there is no reason for these predictions to
be applicable to thermal QCD, and in general the results need to 
be rescaled to be useful at all (see below); this is 
why we refer to the AdS/CFT limit as a ``holographic model''. 
Nevertheless, 
they offer useful qualitative insight into the structures that may
be expected at small $\omega$ and $k$ in an interacting system.

%%%%%%%%%%%%%%%%%%%%%%%%%%%%% SECTION %%%%%%%%%%%%%%%%%%%%%%%%%%%%%%%%%%%%
%
\section{Polynomial interpolation}
\la{se:interp}

As alluded to in \se\ref{ss:nlo}, 
we expect the perturbatively determined $\rho^{ }_\rmii{V}$
to be least precise at {\em small} frequencies. 
For instance deep 
in the spacelike domain ($\ko \ll  k$) only the 
LO result is known (cf.\ \eq\nr{free}), 
but we have argued in \se\ref{ss:hydro} that the true behaviour 
is qualitatively different, at least for very small $k$. Close
to the light cone (for $|M| \lsim g T$), NLO corrections are known, 
but they are only suppressed by $\rmO(g)$ so the
weak-coupling expansion might not converge well. 
In contrast, we may assume that the regime
of large frequencies, known up to $\rmO(g^2)$ 
for $M \gsim \pi T$ and up to $\rmO(g^8)$
for $M \gg \pi T$, is better under control. 

It is an interesting question whether the spectral function
needs to be analytic across the light cone.\footnote{%
 This discussion concerns the 
 infinite-volume limit. 
 } 
At zero temperature this is not the case:
$\rho^{ }_\rmii{V}$ vanishes identically 
in the spacelike domain. However, in an interacting system the spectral
function gets generally smoothened by a temperature. Physical arguments
in favour of smoothness at the NLO level have been presented in 
ref.~\cite{sum2}, and this is also the case in the concrete NLO
computation~\cite{gm} as well as in the non-perturbative frameworks 
discussed in \ses\ref{ss:hydro} and \ref{ss:ads}. In the following, 
we assume $\rho_\rmii{V}^{ }$ to be a smooth function across the light cone,
and represent it through a polynomial interpolation on both sides. 

Let $\omega_0$ lie in the time-like domain, for instance
$\omega_0 \simeq \sqrt{k^2 + (\pi T)^2}$. We introduce 
a polynomial starting with a linear behaviour at $\omega \ll T$ and 
attaching to the known $\rho^{ }_\rmii{V}$ continuously and with a 
continuous first derivative at
$ \ko = \omega_0
$.
Defining
\be
% \rho_\rmii{V}' (0,\vec{k}) \; \equiv \; \alpha
% \;, \quad
 \rho^{ }_\rmii{V} (\omega_0,\vec{k}) \; \equiv \; \beta 
 \;, \quad
 \rho_\rmii{V}' (\omega_0,\vec{k}) \; \equiv \; \gamma 
 \;, \la{match}
\ee
where the dimension of 
$\beta$ is $T^2$ and that 
of $\gamma$ is $T$, 
a general $(5 + 2n_\rmi{max})^\rmi{th}$ order 
polynomial proceeding in odd powers
of $\omega$ and satisfying these boundary values can be expressed as
\be
 \rho^{ }_\rmi{fit} \; \equiv \;
% \alpha\, \omega\, \biggl( 1 - \frac{\omega^2}{\omega_0^2} \biggr)^2 
  \frac{\beta\, \omega^3}{2 \omega_0^3}
 \,\biggl( 5 - \frac{3 \omega^2}{\omega_0^2} \biggr)
 - \frac{\gamma\, \omega^3}{2 \omega_0^2}
 \,\biggl( 1 - \frac{\omega^2}{\omega_0^2} \biggr) 
 + \sum_{n\ge 0}^{n_\rmi{max}}  
  \frac{\delta^{ }_n \omega^{1+2 n}}{\omega_0^{1+2n}}
  \biggl( 1 - \frac{\omega^2}{\omega_0^2} \biggr)^2
% \;, \quad
% 0 < \omega < \omega_0
 \;. \la{fit}
\ee
We treat $\beta$ and $\gamma$ as known from perturbation theory
through the matching in \eq\nr{match}. 
For $n_\rmi{max}=0$ there is only one free parameter in the 5th order
polynomial, given by the slope at origin 
($\alpha \equiv \delta^{ }_0/\omega^{ }_0$), and more 
generally there are $n_\rmi{max} + 1$ free parameters
($\alpha, \delta^{ }_1, ...$). 
For $\omega > \omega^{ }_0$, a perturbative result 
is used (its details are explained in footnote~\ref{fn:kn0}). 

%%%%%%%%%%%%%%%%%%%%%%%%%%%%% SECTION %%%%%%%%%%%%%%%%%%%%%%%%%%%%%%%%%%%%
%
\section{Lattice analysis}
\la{se:latt}

%%%%%%%%%%%%%%%%%%%%%%%%%%%%% SUBSECTION %%%%%%%%%%%%%%%%%%%%%%%%%%%%%%%%%
%
\subsection{Observable and parameters}

In continuum notation, the imaginary-time observable measured
on the lattice reads
\be
 G_\rmii{V}^{ }(\tau,\vec{k}) 
 \; \equiv \;
 \int_{\vec{x}}  
  e^{- i \vec{k}\cdot\vec{x}}
 \Big\langle 
     V^{i}_{ } (\tau,\vec{x}) \, 
     V^{i}_{ } (0) 
 -
     V^{0}_{ } (\tau,\vec{x}) \, 
     V^{0}_{ } (0) 
 \Big\rangle^{ }_\rmi{c}
 \;. \la{GV}
\ee
In order to minimize discretization effects, the momentum is taken to 
point along one of the lattice axes. In a finite-size box momenta are
of the type $k = 2 \pi n /(a N_s)$, where $a$ is the lattice spacing
and $n$ is an integer; 
given that $a N_\tau = 1/T$, we thus consider
\be
 k = 2\pi n T\times  \frac{N_\tau}{N_s}
 \;, \la{k_n}
\ee
where $N_\tau$ and $N_s$ are the temporal and spatial lattice extents, 
respectively. 

The set of lattice simulations considered in the present study
is listed in table~\ref{table:params}. The aspect
ratio was kept fixed at $N_s / N_\tau = 3$ for $T = 1.1\Tc$ 
and at $N_s / N_\tau = 24/7$ for $T = 1.3\Tc$. 
Employing $n\in\{1,2,3\}$ in \eq\nr{k_n} 
the momenta were thus 
$k/T \in\{2.094,4.189,6.283\}$
 and
$k/T \in\{1.833,3.665,5.498 \}$
for $T = 1.1\Tc$ and 
 $T = 1.3\Tc$, respectively.
In order to consider smaller momenta, relevant for 
reaching the hydrodynamic regime, larger $N_s$ should 
be simulated. On the other hand, for the phenomenology of 
photon production, these values appear to be quite reasonable. 

Our measurements were separated by 500 sweeps, each consisting
of 1 heatbath and 4 overrelaxation updates. However, the large
values $\beta_0 \gsim 7.2$ needed imply that topological degrees of freedom
do not thermalize properly even with this much updating, 
so that in general errors may be 
underestimated~\cite{slow}. Given that at $T > \Tc$ 
the physical value of the topological susceptibility is small and
that our observables should not couple much to the slow modes, 
we do not expect to be significantly affected by this 
problem, even if in practice our simulations are frozen to the 
trivial topological sector. 

%%%%%%%%%%%%%%%%%%%%% TABLE %%%%%%%%%%%%%%%%%%%%%%%%%%%%%%%%%%%%%
%
\begin{table}[t]

\small{
\begin{center}
\begin{tabular}{lllllll}
 \hline \\[-3mm]
 $\beta^{ }_0$ &
 $\Ns^3 \times \Nt$ &
 confs & 
 $ T \sqrt{\tO}{ }^{\rmi{ }} $ & 
 $ \left. T / \Tc \right|^\rmi{ }_{\tO} $  & 
 $ T \rO $ & 
 $ \left. T / \Tc \right|^{ }_{\rO} $  \\[3mm]
 \hline 
  7.192  & $96^3 \times 32$  & 314 & 0.2796 & 1.12 
    & 0.816 & 1.09 \\ 
  7.544  & $144^3 \times 48$ & 358 & 0.2843 & 1.14 
    & 0.817 & 1.10 \\ 
  7.793  & $192^3 \times 64$ & 242 & 0.2862 & 1.15 
    & 0.813 & 1.09 \\ \hline
  7.192  & $96^3 \times 28$  & 232 & 0.3195 & 1.28 
    & 0.933 & 1.25 \\ 
  7.544  & $144^3 \times 42$ & 417 & 0.3249 & 1.31 
    & 0.934 & 1.25 \\ 
  7.793  & $192^3 \times 56$ & 273 & 0.3271 & 1.31 
    & 0.929 & 1.25 \\ 
 \hline 
\end{tabular} 
\end{center}
}

\vspace*{3mm}

\caption[a]{\small
  The lattices included in the current analysis, with $\beta_0$
  denoting the coefficient of the Wilson plaquette term. Simulations
  are carried out within quenched SU(3) gauge theory.  
  Conversions to units of $\tO$~\cite{t0}, $\rO$~\cite{r0}
  and $\Tc$ are based on ref.~\cite{betac}. 
  In a separate set of simulations at a somewhat higher
  temperature~\cite{cond3}, 
  spatial volume dependence has been verified
  to be within statistical errors. 
 }
\label{table:params}
\end{table}
%
%%%%%%%%%%%%%%%%%%%%%%%%%%%%%%%%%%%%%%%%%%%%%%%%%%%%%%%%%%%%%%%%%%%%%

%%%%%%%%%%%%%%%%%%%%%%%%%%%%% SUBSECTION %%%%%%%%%%%%%%%%%%%%%%%%%%%%%%%%%
%
\subsection{Continuum extrapolation}

For the lattice analysis we employed a local discretization of 
the vector current, with non-perturbatively 
clover-improved Wilson fermions~\cite{clover1,clover2}.  
As discussed in \se\ref{ss:basic}, only the connected quark 
contraction needs to be evaluated for the observable that
we are interested in. The general techniques of the lattice
analysis have been discussed in ref.~\cite{cond3}, and  
the ensemble employed 
for our numerical investigation in ref.~\cite{setup}. 

We carry out a continuum extrapolation
for the ratios $G^{ }_\rmii{V}(\tau,\vec{k}) T^2 / 
[\chi^{ }_\rmi{q} G^{ }_{\rmii{V}, \rmi{free}}(\tau,\vec{0}) ]  $, 
where $\chi^{ }_\rmi{q}$ is the quark number susceptibility and 
\be
 G^{ }_{\rmii{V}, \rmi{free}}(\tau,\vec{0})
 \; \equiv \; 
  6 T^3 \biggl[ 
  \pi (1-2\tau T) \frac{1 + \cos^2(2\pi\tau T)}{\sin^3(2\pi \tau T)}
 + 
 \frac{2 \cos(2\pi \tau T)}{\sin^2(2\pi \tau T)}
  \biggr]
 \;. \la{GVfree}
\ee
Normalization by $\chi^{ }_\rmi{q}$
removes the renormalization factors associated with our local discretization
of the vector current, and normalization through 
$
  G^{ }_{\rmii{V}, \rmi{free}}
$
hides the short-distance growth of the imaginary-time correlator. 
$\rmO(a)$ improvement permits for a 
continuum extrapolation quadratic in $1/N^{ }_\tau$.
More details can be found in ref.~\cite{setup}.
With this approach a continuum extrapolation could be carried out at
    $\tau T \ge 0.18$ for $T=1.1T_c$
   and at $\tau T \ge 0.22$ for $T=1.3T_c$.
These are the distances included in the subsequent analysis. 
A bootstrap sample was generated for the continuum extrapolated
results, which was used for estimating the statistical errors of
our final observables. In a separate set of continuum extrapolations, 
the susceptibilities were determined through
a quadratic fit, yielding
$\chi^{ }_\rmi{q} =  0.857(16) T^2$ at $T = 1.1 \Tc$ and 
$\chi^{ }_\rmi{q} =  0.897(17) T^2$ at $T = 1.3 \Tc$~\cite{setup}.

%%%%%%%%%%%%%%%%%%%%%%%%%%%%% SECTION %%%%%%%%%%%%%%%%%%%%%%%%%%%%%%%%%%%%
%
\section{Fit results}
\la{se:fit}

%%%%%%%%%%%%%%%%%%%%%%%%%%%%%%%%% FIGURE %%%%%%%%%%%%%%%%%%%%%%%%%%%%%%%%%
\begin{figure}[t]

\hspace*{-0.1cm}
\centerline{%
 \epsfysize=7.5cm\epsfbox{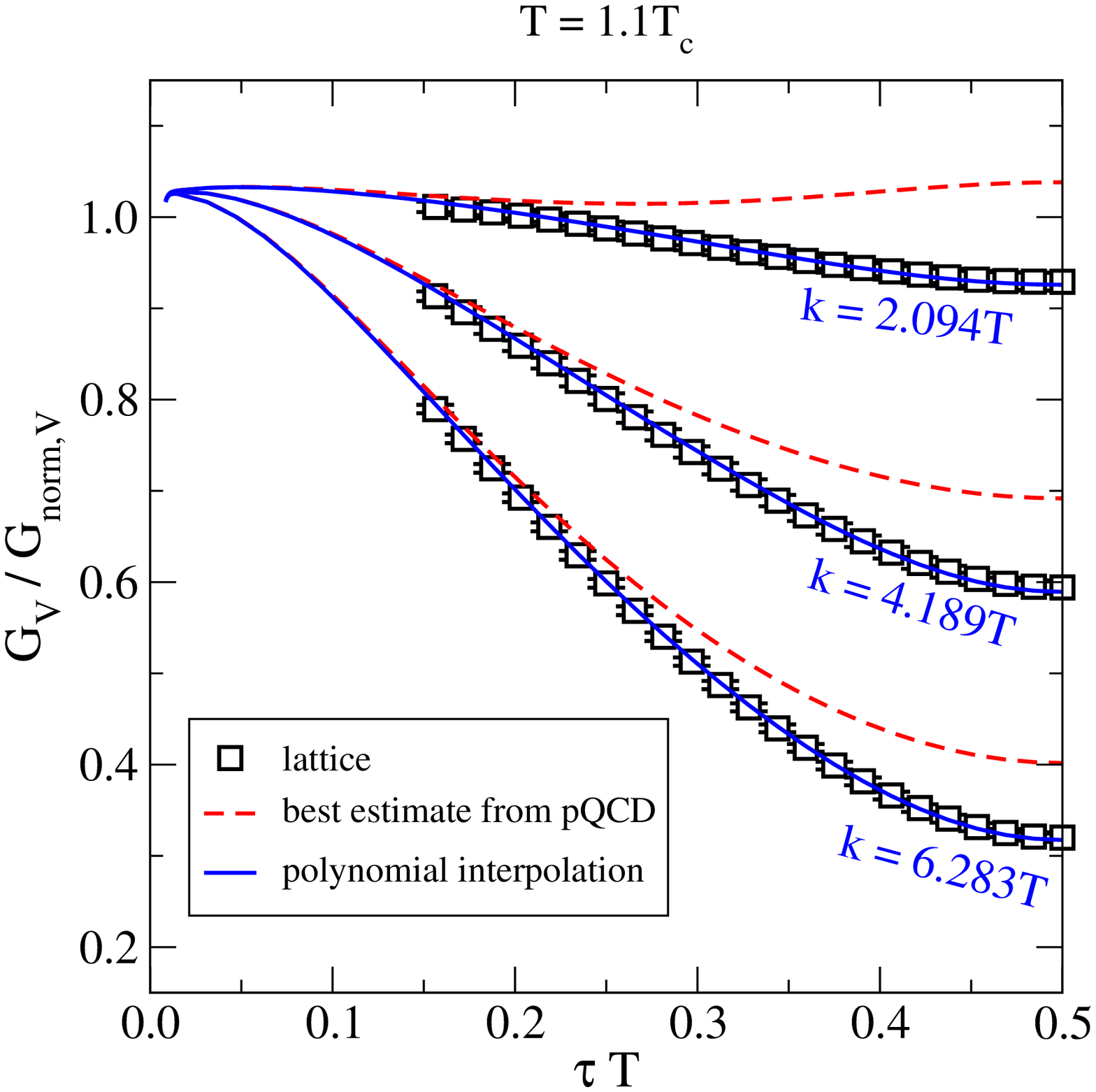}%
 \hspace{0.1cm}
 \epsfysize=7.5cm\epsfbox{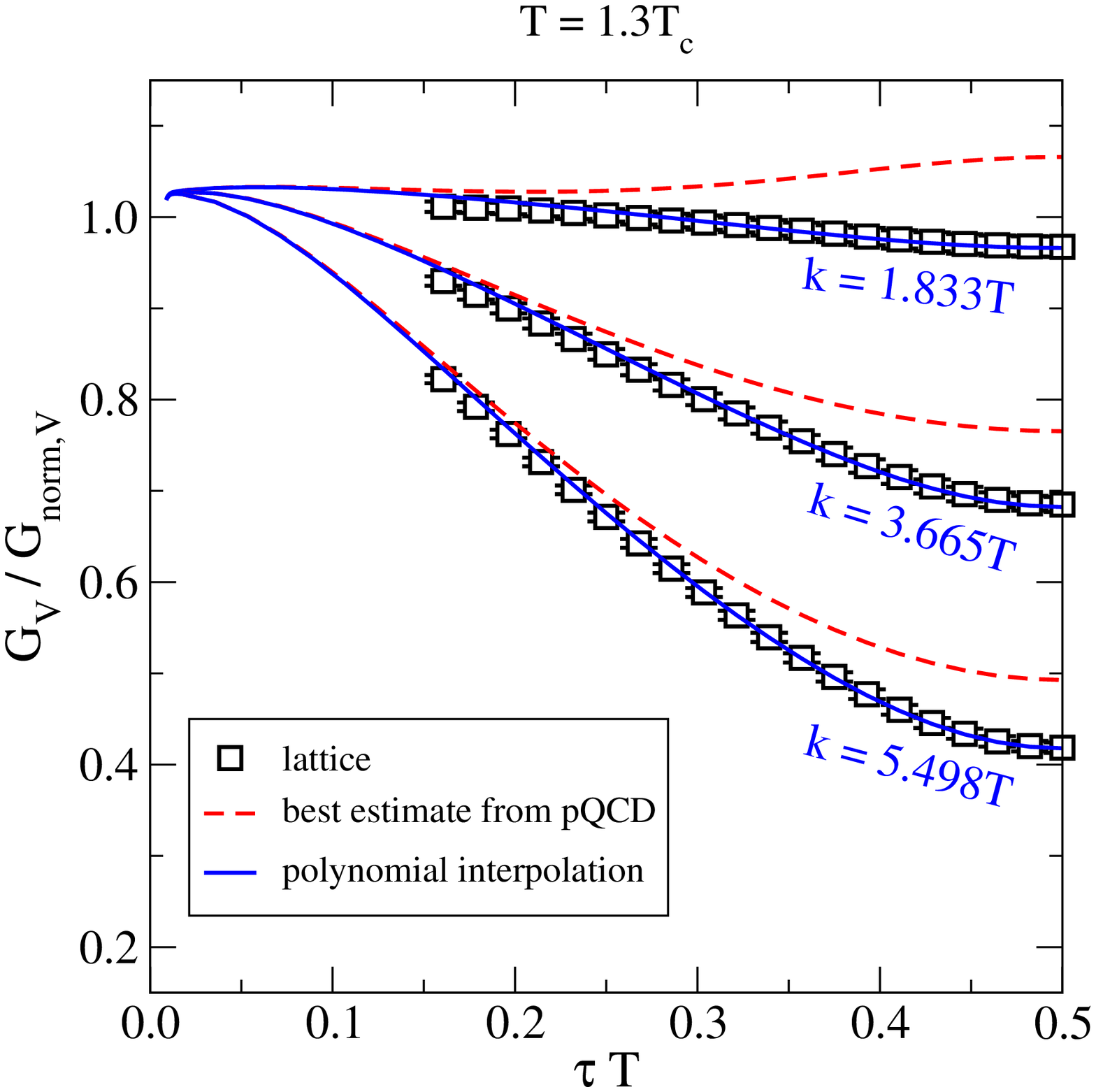}
}

\caption[a]{\small
 Fitted imaginary-time correlators at non-zero momenta. 
 The ``best estimate from pQCD'' (perturbative QCD)
 is based on refs.~\cite{gm,lpm,dilepton}, and has 
 been constructed as explained in footnote~\ref{fn:kn0}.
 ``Polynomial interpolations'' correspond
 to $n^{ }_\rmi{max} = 0$, but similarly good fits 
 are obtained for $n^{ }_\rmi{max} = 1$.
}

\la{fig:GV}
\end{figure}
%%%%%%%%%%%%%%%%%%%%%%%%%%%%%%%%%%%%%%%%%%%%%%%%%%%%%%%%%%%%%%%%%%%%%%%%%%%

%%%%%%%%%%%%%%%%%%%%%%%%%%%%%%%%% FIGURE %%%%%%%%%%%%%%%%%%%%%%%%%%%%%%%%%
\begin{figure}[t]

\hspace*{-0.1cm}
\centerline{%
 \epsfysize=7.5cm\epsfbox{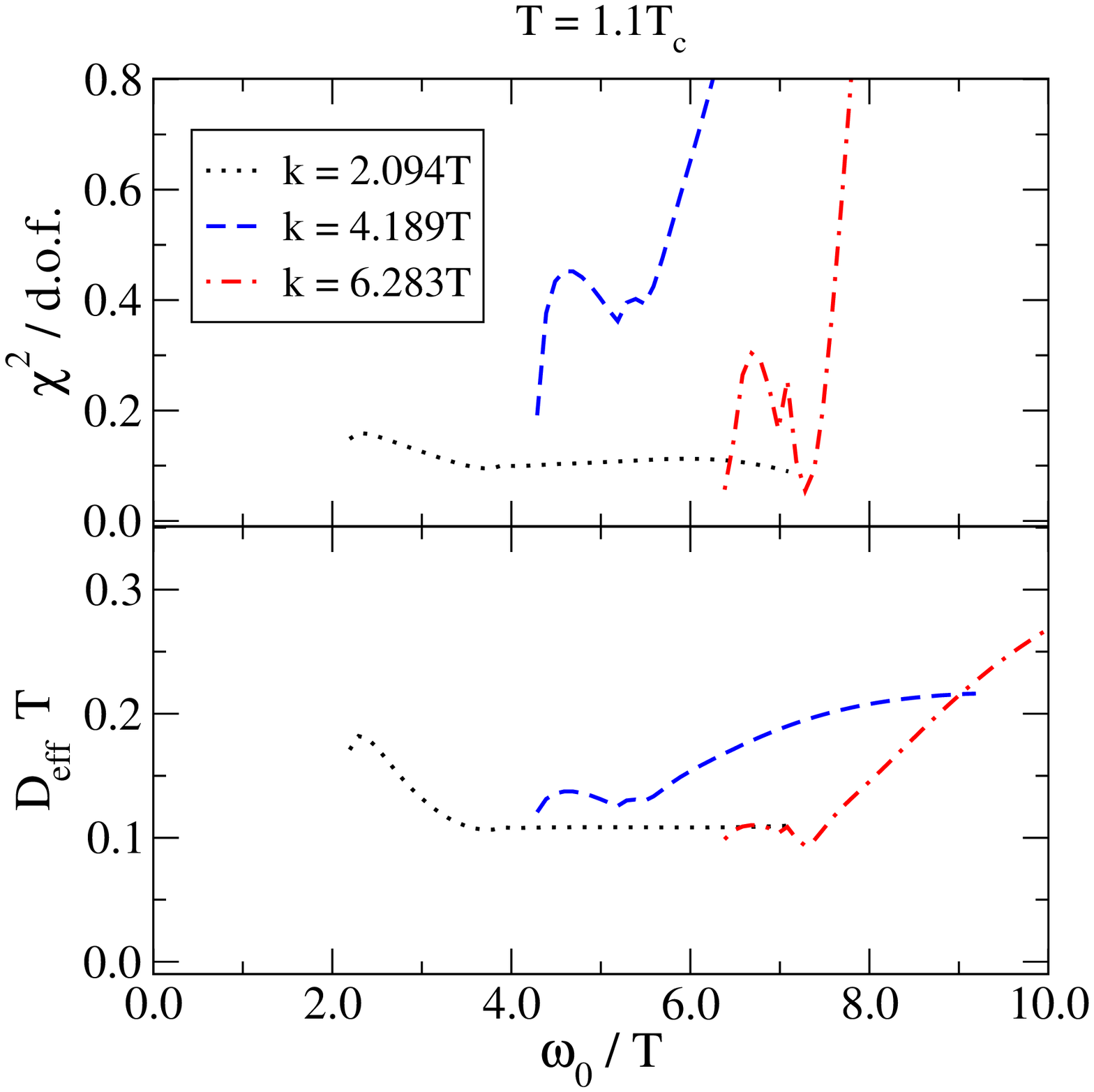}%
 \hspace{0.1cm}
 \epsfysize=7.5cm\epsfbox{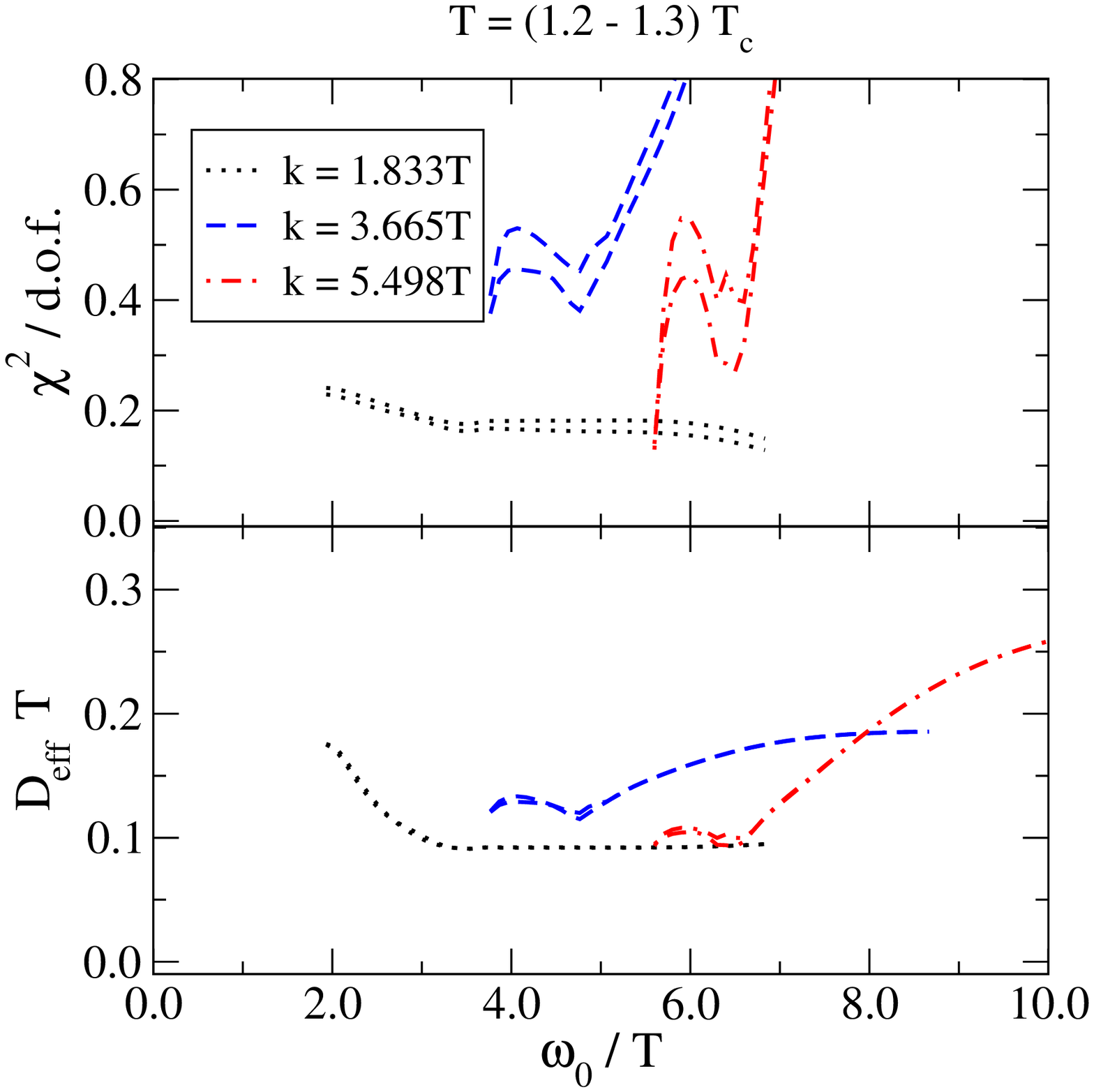}
}

\caption[a]{\small
 We show $\chi^2/\mbox{d.o.f.}$ (top) 
 and $D^{ }_\rmi{eff} T$ (bottom; cf.\ \eq\nr{Deff})
 as a function of the matching point~$\omega_0$ for $n_\rmi{max} = 0$.
 In the right panel, the upper curves are for $T=1.2\Tc$ and 
 the lower curves for $T=1.3\Tc$ on the perturbative side
 (the lattice data is fixed but it is not known precisely 
 to which temperature it corresponds, cf.\ table~\ref{table:params}). 
 A local minimum of $\chi^2/\mbox{d.o.f.}$ is generally found
 close to the point where $\omega_0 = \sqrt{k^2 + (\pi T)^2}$; 
 it is very shallow for the smallest $k$.
}

\la{fig:chisq}
\end{figure}
%%%%%%%%%%%%%%%%%%%%%%%%%%%%%%%%%%%%%%%%%%%%%%%%%%%%%%%%%%%%%%%%%%%%%%%%%%%

%%%%%%%%%%%%%%%%%%%%%%%%%%%%%%%%% FIGURE %%%%%%%%%%%%%%%%%%%%%%%%%%%%%%%%%
\begin{figure}[t]

\hspace*{-0.1cm}
\centerline{%
 \epsfysize=7.5cm\epsfbox{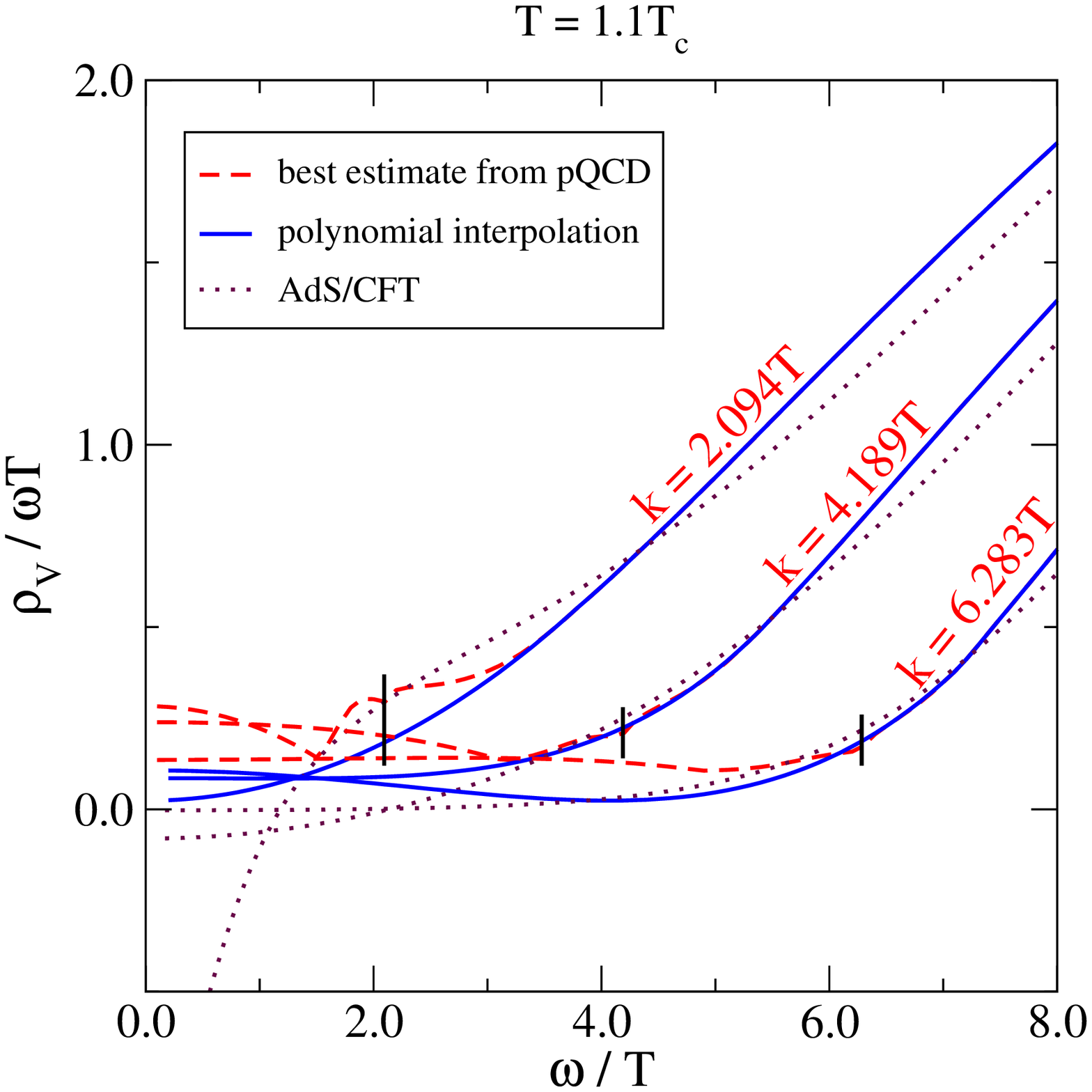}%
 \hspace{0.1cm}
 \epsfysize=7.5cm\epsfbox{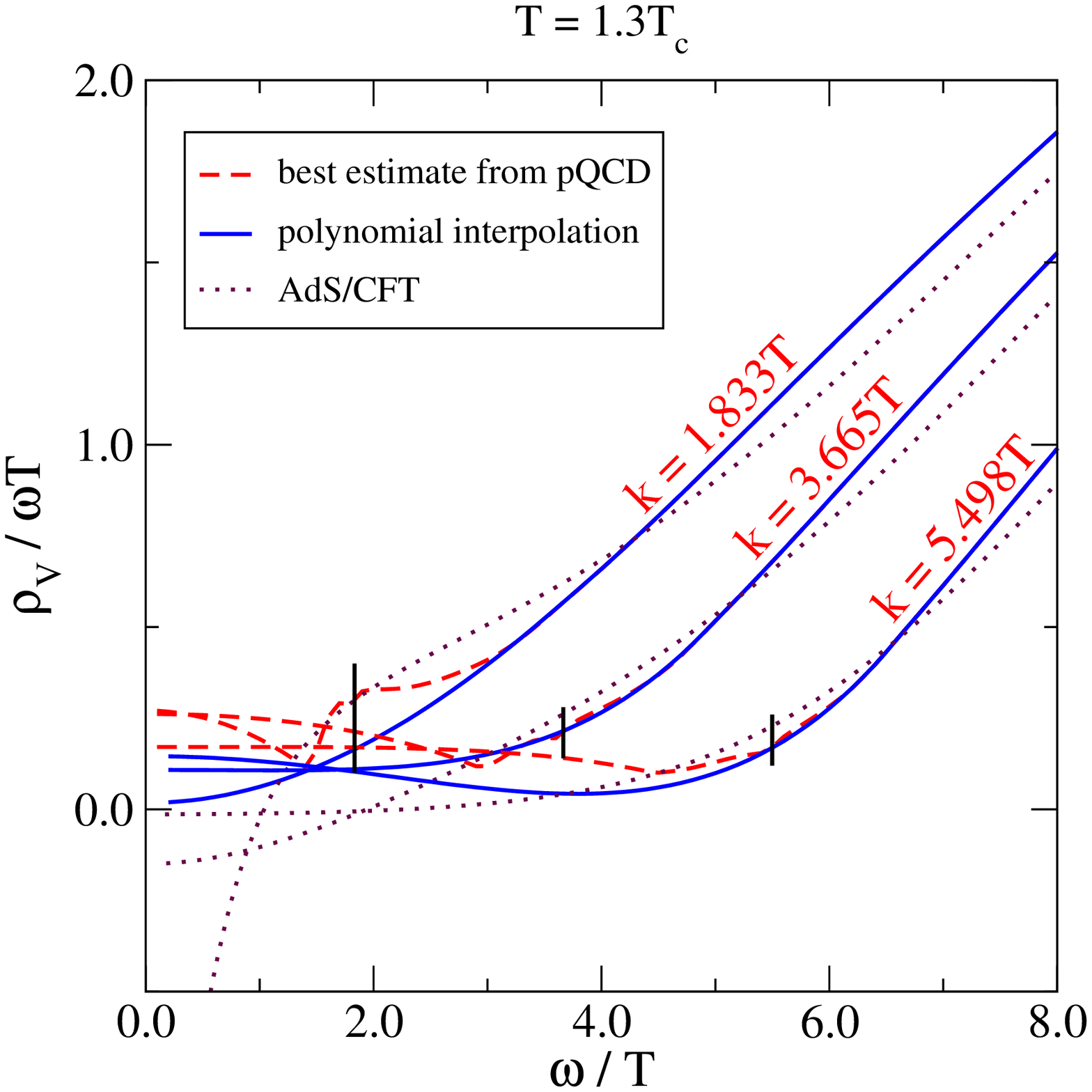}
}

\caption[a]{\small
 The spectral functions corresponding to \fig\ref{fig:GV} ($n_\rmi{max} = 0$). 
 The vertical bars locate the light cone. 
 The ``best estimate from pQCD''
 is based on refs.~\cite{gm,lpm,dilepton}, and has 
 been constructed as explained in footnote~\ref{fn:kn0}.
 The AdS/CFT result comes from ref.~\cite{ads2}, and has been rescaled
 to agree with the non-interacting QCD result at large $\omega/T$.
 (This rescaling choice is rather arbitrary.)
}

\la{fig:rho_kn0}
\end{figure}
%%%%%%%%%%%%%%%%%%%%%%%%%%%%%%%%%%%%%%%%%%%%%%%%%%%%%%%%%%%%%%%%%%%%%%%%%%%

%%%%%%%%%%%%%%%%%%%%%%%%%%%%%%%%% FIGURE %%%%%%%%%%%%%%%%%%%%%%%%%%%%%%%%%
\begin{figure}[t]

\hspace*{-1.1cm}
\centerline{%
 \epsfysize=7.5cm\epsfbox{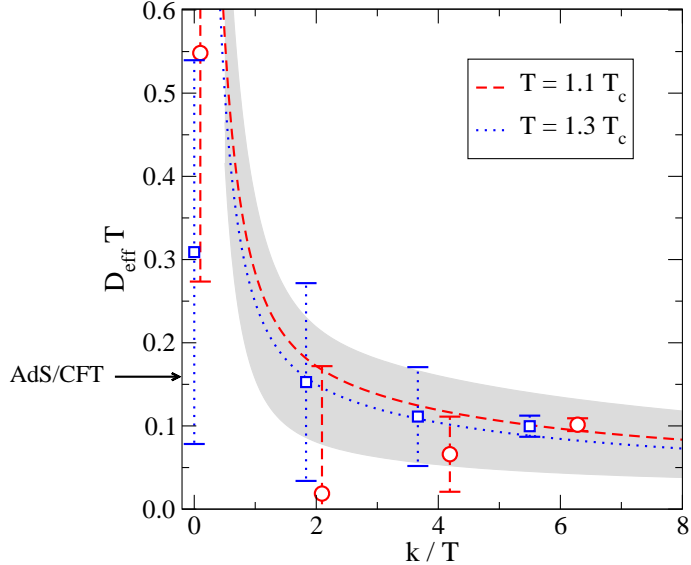}%
% \hspace{0.1cm}
% \epsfysize=7.5cm\epsfbox{ }
}

\caption[a]{\small
 Lattice results for $D_\rmi{eff}$ defined in \eq\nr{Deff} (data points), 
 compared with the NLO perturbative prediction 
 from ref.~\cite{gm} (continuous curves).
 The lattice errors have been 
 obtained by carrying out fits with $n_\rmi{max} = 1$ to the 
 bootstrap ensemble. 
 The data points at $k = 0$ (cf.\ appendix~A) have been 
 slightly displaced for better 
 visibility. For comparison note that 
 the heavy-quark diffusion coefficient, determined with 
 different methods, has been estimated as $DT \sim 0.6 ... 1.1$
 at $T\sim 1.5\Tc$~\cite{kappa},  
 and the light-quark value as 
 $DT \sim 0.2 ... 0.8$ at $T = 1.1\Tc$ and
 $DT \sim 0.2 ... 0.5$ at $T = 1.3\Tc$~\cite{setup}.
 The predictions of ref.~\cite{gm} are only reliable for $k \gg gT$, 
 but LO perturbative values 
 at $k=0$ can be obtained by dividing the results  
 of ref.~\cite{amy2} through the lattice susceptibility 
 according to \eq\nr{Kubo_D}, yielding $DT \approx 2.9$
 at $T = 1.1\Tc$ and $DT \approx 3.1$ at $T = 1.3\Tc$.
 The AdS/CFT value is $DT = 1/(2\pi)$~\cite{ads1}.
}

\la{fig:Deff}
\end{figure}
%%%%%%%%%%%%%%%%%%%%%%%%%%%%%%%%%%%%%%%%%%%%%%%%%%%%%%%%%%%%%%%%%%%%%%%%%%%

%%%%%%%%%%%%%%%%%%%%% TABLE %%%%%%%%%%%%%%%%%%%%%%%%%%%%%%%%%%%%%
%
\begin{table}[t]

\small{
\begin{center}
\begin{tabular}{lllllll}
 \hline \\[-3mm]
 $T/\Tc$ &
 $k/T$ &
 $\alpha / T$ & 
 $\beta / T^2$ & 
 $\gamma / T$  & 
 $ \left. T D_\rmi{eff} \right|^{ }_\rmi{$n_\rmii{max} = 0$} $ & 
 $ \left. T D_\rmi{eff} \right|^{ }_\rmi{$n_\rmii{max} = 1$} $
 \\[2mm]  \hline
 1.1 & 2.094 & 0.028(15) & 2.072 & 1.611  & 0.108(4) & 0.019(153) \\
     & 4.189 & 0.091(8) & 2.325 & 1.963   & 0.130(1) & 0.066(45) \\
     & 6.283 & 0.105(4) & 2.498 & 2.331   & 0.109(1) & 0.102(8)
 \\ \hline
 1.3 & 1.833 & 0.024(17) & 2.038 & 1.558  & 0.093(5) & 0.153(119) \\
     & 3.665 & 0.112(10) & 2.229 & 1.984  & 0.119(1) & 0.111(59) \\
     & 5.498 & 0.141(6)  & 2.367 & 2.438  & 0.094(1) & 0.097(13)
 \\  \hline 
\end{tabular} 
\end{center}
}

\vspace*{3mm}

\caption[a]{\small
  Fit results for the coefficients in \eq\nr{fit}, 
  with $\alpha = \delta^{ }_0 / \omega^{ }_0$, and 
  for the effective diffusion coefficient $D_\rmi{eff}$ of \eq\nr{Deff}, 
  from fits with $n_\rmi{max} = 0$. 
  For $D_\rmi{eff}$ the results from the bootstrap analysis
  with $n_\rmi{max} = 1$
  are also shown; the latter constitute our final results 
  and are illustrated in \fig\ref{fig:Deff}. 
 }
\label{table:coeffs}
\end{table}
%
%%%%%%%%%%%%%%%%%%%%%%%%%%%%%%%%%%%%%%%%%%%%%%%%%%%%%%%%%%%%%%%%%%%%%

Having discussed the spectral function on one side
(\se\ref{se:interp}) and the 
imaginary-time correlator on the other
(\se\ref{se:latt}), the remaining task is
to compare the two. 
The relation is given by
\be
 G^{ }_\rmii{V}(\tau,\vec{k})
 = 
 \int_0^\infty \! \frac{{\rm d}\omega}{\pi}
 \, \rho^{ }_\rmii{V}(\omega,\vec{k}) \, 
 \frac{\cosh [\omega (\frac{\beta }{2}-\tau) ] }
 {\sinh [ \frac{\omega \beta }{2} ] }
 \;, \quad
 \beta \equiv \frac{1}{T}
 \;. 
 \la{relation}
\ee
Inserting
into \eq\nr{relation} the best available perturbative estimate for 
$\rho^{ }_\rmii{V}$, based on an interpolation between 
the results of refs.~\cite{gm,lpm,dilepton},\footnote{%
 The data is available through ref.~\cite{dilepton-lattice}.
 More precisely, for very large time-like frequencies it is given by 
 the large-$M$ results of ref.~\cite{dilepton} which go over
 into the N$^4$LO vacuum result for 
 $\omega \gg \pi T$~\cite{ope,kit_ns,kit_si}.
 For $\omega\lsim 10 T$
 it is given by the interpolation
 of the large-$M$ result and 
 the LO LPM-resummed small-$M$ result, as 
 presented in ref.~\cite{lpm}, 
 summed together with the NLO small-$M$ result of
 ref.~\cite{gm} (switched off exponentially with growing $M$ to avoid
 OPE-violating contributions \cite{ope} proportional to $T^2$). 
 In this way,
 the value at the real photon point $\omega=k$ agrees with the NLO
 photon calculation~\cite{ak}.
 In the space-like region the spectral function is the largest between
 the Born one with vacuum corrections \cite{dilepton} and the 
 NLO small-$M$ result~\cite{gm}. In practice, this implies
 that at the smallest $\omega$ we have the Born-like spectral function,
 whereas close to the light-cone we have the small-$M$ one,
 ensuring continuity across the light-cone. \la{fn:kn0}}
a visible discrepancy is observed between
the perturbative and lattice results at $\tau T \gsim 0.3$ 
(cf.\ \fig\ref{fig:GV}). 
In general the lattice results are {\em below} the perturbative
ones. The goal now is to test whether the discrepancy could be 
explained by modifications of $\rho^{ }_\rmii{V}$ in the domain
of small frequencies, as explained in \se\ref{se:interp}.

With the ansatz
of \eq\nr{fit}, a good representation of the data can 
indeed be obtained. 
% (we omit distances $\tau T \lsim 0.18$ from the fit). 
This is illustrated in \fig\ref{fig:GV} and more quantitatively 
in \fig\ref{fig:chisq}, which shows the dependence of
$\chi^2$ on the matching point $\omega_0$. 
In the following, we fix $\omega_0 = \sqrt{k^2 + (\pi T)^2}$, 
which is close to the local minimum of $\chi^2$.
A small $\chi^2$ could also be obtained with $\omega_0 = k$, 
where the curves start, but we prefer to use the minimum 
that is deeper in the perturbative domain, because then 
we have more reasons to trust the perturbative prediction.

The corresponding results for 
the spectral function are illustrated
in \fig\ref{fig:rho_kn0}. 
Barring the possibility of large non-perturbative effects
at $M  \gsim \pi T$, it appears plausible from \fig\ref{fig:rho_kn0}
that the pQCD spectral functions have too much
weight in the spacelike domain. This is in qualitative agreement
with the discussion in \ses\ref{ss:hydro} and \ref{ss:ads}, and 
suggests the gradual onset of hydrodynamics-like behaviour.
That the fit lies below the perturbative curves
at $k \lsim 3 T$ is also consistent with the expectation that
the diffusion coefficient $D$ of a strongly coupled system should be smaller 
than the result of a leading-order weak-coupling analysis~\cite{amy2}.

The value of the spectral function at the photon point, normalized
as  $\rho^{ }_\rmii{V}(k,\vec{k})T/(2 \chi^{ }_\rmi{q} k)$, is shown in 
\fig\ref{fig:chisq} (lower panels) 
and in \fig\ref{fig:Deff}. More precisely, in order to 
accommodate data both at $k=0$ and at $k>0$, we define
\be
 D_\rmi{eff}(k)  \;\equiv\;
 \left\{
 \begin{array}{ll}
    \displaystyle \frac{\rho^{ }_\rmii{V}(k,\vec{k})}{2 \chi^{ }_\rmi{q} k} 
 & \;, \quad k > 0 \\
    \displaystyle 
    \lim_{\omega\to 0^+}
 \frac{\rho_{ }^{ii}(\omega,\vec{0})}{3 \chi^{ }_\rmi{q} \omega}
 & \;, \quad k = 0 
 \end{array}
 \right.
 \;. \la{Deff}
\ee
According to \eqs\nr{Kubo_D} and \nr{hydro}, 
$\lim_{k\to 0} D_\rmi{eff}(k) = D$. Even though the evidence for 
a continuous behaviour is not overwhelming in \fig\ref{fig:Deff} 
due to the large 
systematic uncertainties at small $k \lsim 3 T$, it is not 
excluded either. We recall that according to the 
discussion in \se\ref{ss:ads}, hydrodynamic behaviour is
expected to set in for $k \lsim 1/D$, which according to 
the $k=0$ results in \fig\ref{fig:Deff} roughly
speaking corresponds to $k \lsim 2 T$. 

As already alluded to, 
our analysis contains systematic as well 
as statistical uncertainties. 
In order get an impression about their magnitudes, 
the following tests have been carried out: 

\begin{itemize}

\item
We have tested the dependence of the results on the order of 
the fitted polynomial, parametrized by $n_\rmi{max}$ in \eq\nr{fit}. 
Obviously, given the ill-posed nature of the inversion problem, 
the results are quite sensitive to $n_\rmi{max}$. 
The difference of the results obtained with 
$n_\rmi{max} = 0$ and $n_\rmi{max} = 1$ 
can be employed as one indication of systematic errors, 
cf.\ table~\ref{table:coeffs}.
The resulting errors are of the same order of magnitude but
somewhat smaller than those obtained from the bootstrap sample
with $n_\rmi{max} = 1$, cf.\ table~\ref{table:coeffs} and the 
discussion below.  Therefore we display
the latter as our uncertainties in \fig\ref{fig:Deff}. 
Stable results (i.e.\ results with errors below 100\%)
could only be obtained for $k\gsim 3 T$.

\item
On the lattice side, uncertainties related to scale fixing imply a 
certain uncertainty of the value of $T/\Tc$ simulated, 
cf.\ table~\ref{table:params}. On the perturbative side, there is
an uncertainty from higher orders in the perturbative expansion, which
can partly be estimated through the dependence of the results on 
the renormalization scale. Our experience suggests that 
the latter scale uncertainty (which is a higher-order effect) is 
of a similar magnitude as the former (which is a leading-order
effect but with a smaller variation). We show results from a variation
of the former type in the right panel of \fig\ref{fig:chisq}, concluding
that this uncertainty is negligible compared with 
the dependence on $n_\rmi{max}$.

\item
As mentioned above, our continuum extrapolations were carried out 
for the ratios 
$T^2 G^{ }_\rmii{V}/[ \chi^{ }_\rmi{q}\,G^{ }_{\rmii{V},\rmi{free}}]$, 
and the continuum value of 
$\chi^{ }_\rmi{q}/T^2$ was determined 
through a separate extrapolation. 
For a matching to perturbative results in the ultraviolet regime, 
we need the value of $G^{ }_\rmii{V}/ T^3$. In other words, 
the errors related to the two separate continuum extrapolations 
need to be combined. We have done this 
by fixing $\chi^{ }_\rmi{q}/T^2$ to its central, minimal, and maximal
value within the error band, and repeating the bootstrap 
analysis in each case. The resulting variations of  
$D^{ }_\rmi{eff}T$ are 
subleading compared with systematic uncertainties, 
and can be omitted in practice. 

\item
In \figs\ref{fig:GV} and \ref{fig:Gii}, the errors shown for the 
lattice data correspond to diagonal entries of the covariance matrix. 
However, we have carried out a full-fledged bootstrap analysis. 
Bootstrap samples were used for constructing a covariance matrix in 
the $\tau$-regime where the continuum extrapolation was
judged to be reliable. The inverse of the covariance matrix was 
employed in order to determine the $\chi^2$-value of a fit of
any individual configuration to our 
ansatz. The resulting distribution was
used for obtaining errors for $D^{ }_\rmi{eff}$, shown in 
\fig\ref{fig:Deff}.
The results obtained with $n_\rmi{max} = 0$ and $n_\rmi{max} = 1$ are given 
in table~\ref{table:coeffs}. The errors of the $n_\rmi{max} = 1$ results
encompass in general the central values of the $n_\rmi{max} = 0$ results,  
and constitute our best estimate of uncertainties. 

\end{itemize}

%%%%%%%%%%%%%%%%%%%%%%%%%%%%% SECTION %%%%%%%%%%%%%%%%%%%%%%%%%%%%%%%%%%%%
%
\section{Conclusions}
\la{se:concl}

We have shown how a combination of lattice and perturbative results
allows us to obtain non-trivial information about the vector channel
spectral function close to the photon point. 
The results are conveniently displayed in terms of the 
function $D_\rmi{eff}(k)$, defined in \eq\nr{Deff}.
The observed small difference between the fit and the 
perturbative result at $k\gsim 3 T$, cf.\ \fig\ref{fig:Deff}, 
is consistent with the smallness of the 
NLO correction~\cite{ak,gm}, as well as with indirect crosschecks 
concerning the convergence of the weak-coupling expansion for
light-cone observables at $k\gsim 2\pi T$, based on measuring screening masses
at non-zero Matsubara frequencies~\cite{screening}.

We have demonstrated that, even though
not constrained to do so {\it a priori}, the fit result reproduces   
some qualitative features expected from the soft domain, namely 
a reduced (and possibly even negative) spectral weight in the 
spacelike domain, cf.\ \fig\ref{fig:rho_kn0}. Basically, the 
best fit result lies between the pQCD and 
the strong-coupling AdS/CFT predictions.

As has been illustrated in 
\fig\ref{fig:Deff}, 
measurements at non-zero momenta may  
offer for an alternative way to estimate the diffusion 
coefficient, 
avoiding possible problems 
of the standard approach~\cite{cond1,setup,cond2a,cond3,cond5,cond2b,cond6}
which have to deal with a very narrow transport peak 
at zero momentum~\cite{mr}. However, for a quantitative study, 
much smaller values of $k$ should be reached with controlled errors. 
It would be interesting to test whether the analytic improvement program
of ref.~\cite{ff} could help in this. 
Conceivably, a similar methodology could also be employed for 
estimating other transport coefficients, 
such as the shear viscosity of the QCD plasma. 

Our analysis made use of continuum-extrapolated lattice
data for quenched QCD ($\Nf = 0$). However, 
the qualitative lessons are expected to remain valid also for
unquenched QCD. 

In terms of the quantity $D_\rmi{eff}(k)$ defined in \eq\nr{Deff}
and shown in \fig\ref{fig:Deff}, the physical photon rate from 
\eq\nr{photon} can be expressed as (for $\Nf = 3$)
\be
 \frac{{\rm d}\Gamma_\gamma(\vec{k})}{{\rm d}^3\vec{k}}
  \; = \;
 \frac{2  \alpha_\rmi{em} \chi^{ }_\rmi{q} }{3 \pi^2}
 \, \nB{}(k) D_\rmi{eff}(k)
 \;. \la{photon2}
\ee
Here $\chi_\rmi{q} \lsim T^2$ is a 
light quark number susceptibility, and $\nB{}$ is the Bose distribution. 
The parametrization in \eq\nr{photon2} 
should be useful for phenomenological analyses
as well. In particular, given that $D^{ }_\rmi{eff}$ is a decreasing
function of $k$, the soft photon production 
rate increases at small $k$ even faster  than 
the naive estimate
$
 {{\rm d}\Gamma_\gamma} / {{\rm d}^3\vec{k}}
 \sim 
 \alpha_\rmi{em} T \nB{}(k)
$.

To summarize, the present results support the program of 
implementing pQCD results into
hydrodynamical codes~\cite{hydro1,hydro2,hydro3}. 
The theoretical uncertainties could be as low as 
$\sim 20\%$, 
save for soft $k \lsim 2 T$ where the pQCD results
represent an overestimate (cf.\ \fig\ref{fig:Deff}). 
It is remarkable that
such an overshooting is in apparent qualitative 
agreement with phenomenology~\cite{hydro2}. 
In light of the photon $v_2$ puzzle, 
it would be interesting to extend the investigation down to lower
temperatures, even though this is well justified only  in the 
presence of dynamical quarks and even though at low temperatures
the spectral function ansatz
should include the possibility of vector resonance contributions. 

%%%%%%%%%%%%%%%%%%%%%%%%%%%%% SECTION %%%%%%%%%%%%%%%%%%%%%%%%%%%%%%%%%%%%
%
\section*{Acknowledgments}                                    
               
We thank G.D.~Moore for providing us his codes related to 
refs.~\cite{ads2,mr} and for helpful comments on the manuscript.         
The work has been supported
by the DFG under grant GRK881, 
by the SNF under grant 200020-155935,
by the European Union through HadronPhysics3
(grant 283286) and ITN STRONGnet
(grant 238353), and by the V\"ais\"al\"a Foundation.
Simulations were
performed using JARA-HPC resources at the RWTH Aachen and JSC J\"ulich
(projects JARA0039 and JARA0108), 
the OCuLUS Cluster at the Paderborn Center for Parallel
Computing, and the Bielefeld GPU cluster.

%%%%%%%%%%%%%%%%%%%%%%% APPENDIX %%%%%%%%%%%%%%%%%%%%%%%%%%%%%%%%%%%
%
\appendix
\renewcommand{\thesection}{Appendix~\Alph{section}}
\renewcommand{\thesubsection}{\Alph{section}.\arabic{subsection}}
\renewcommand{\theequation}{\Alph{section}.\arabic{equation}}

%%%%%%%%%%%%%%%%%%%%%%%%%%%% SECTION %%%%%%%%%%%%%%%%%%%%%%%%%%%%%%%%
%
\section{Results for zero momentum}

%%%%%%%%%%%%%%%%%%%%%%%%%%%%%%%%% FIGURE %%%%%%%%%%%%%%%%%%%%%%%%%%%%%%%%%
\begin{figure}[t]

\hspace*{-0.1cm}
\centerline{%
 \epsfysize=7.5cm\epsfbox{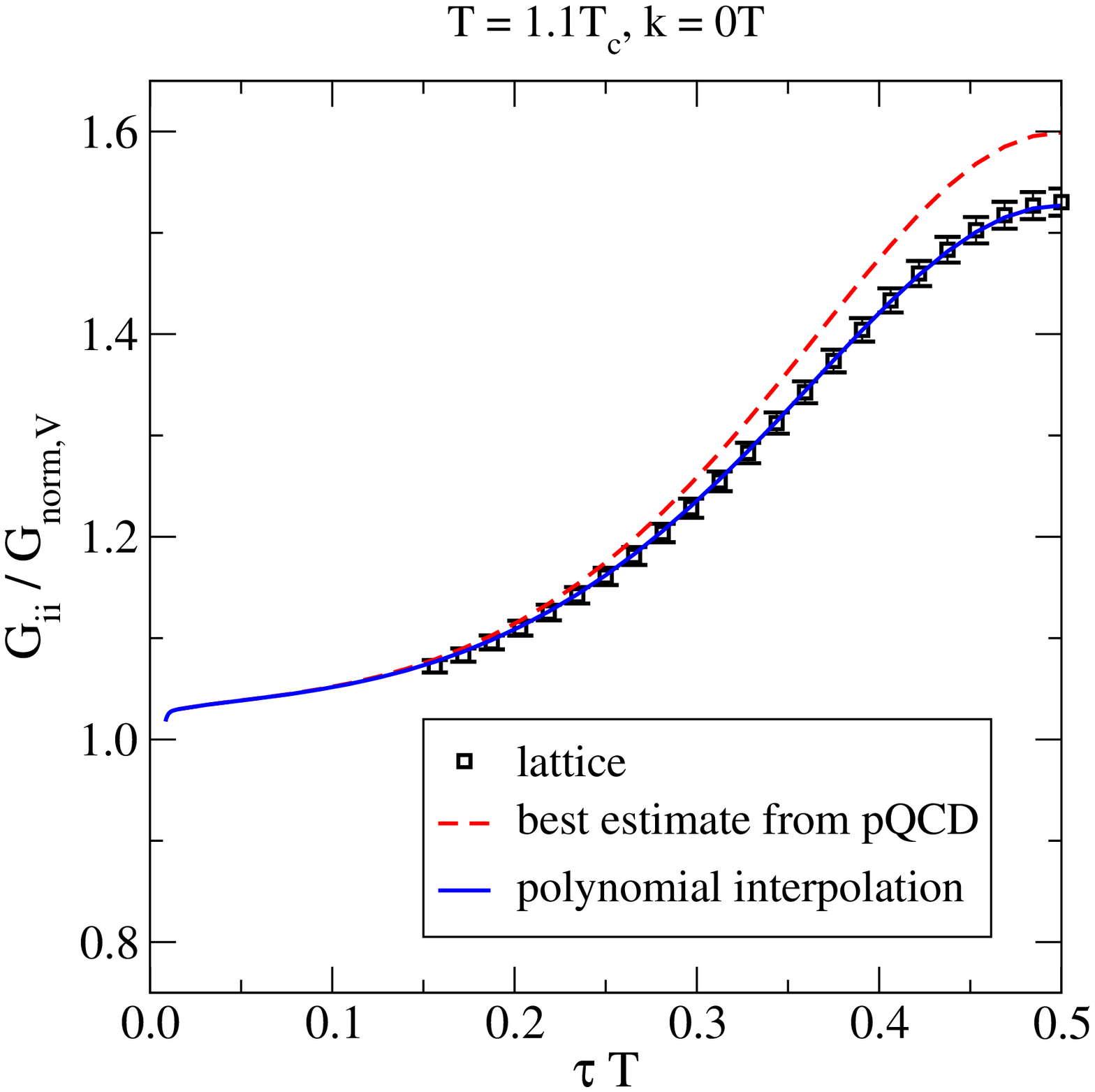}%
 \hspace{0.1cm}
 \epsfysize=7.5cm\epsfbox{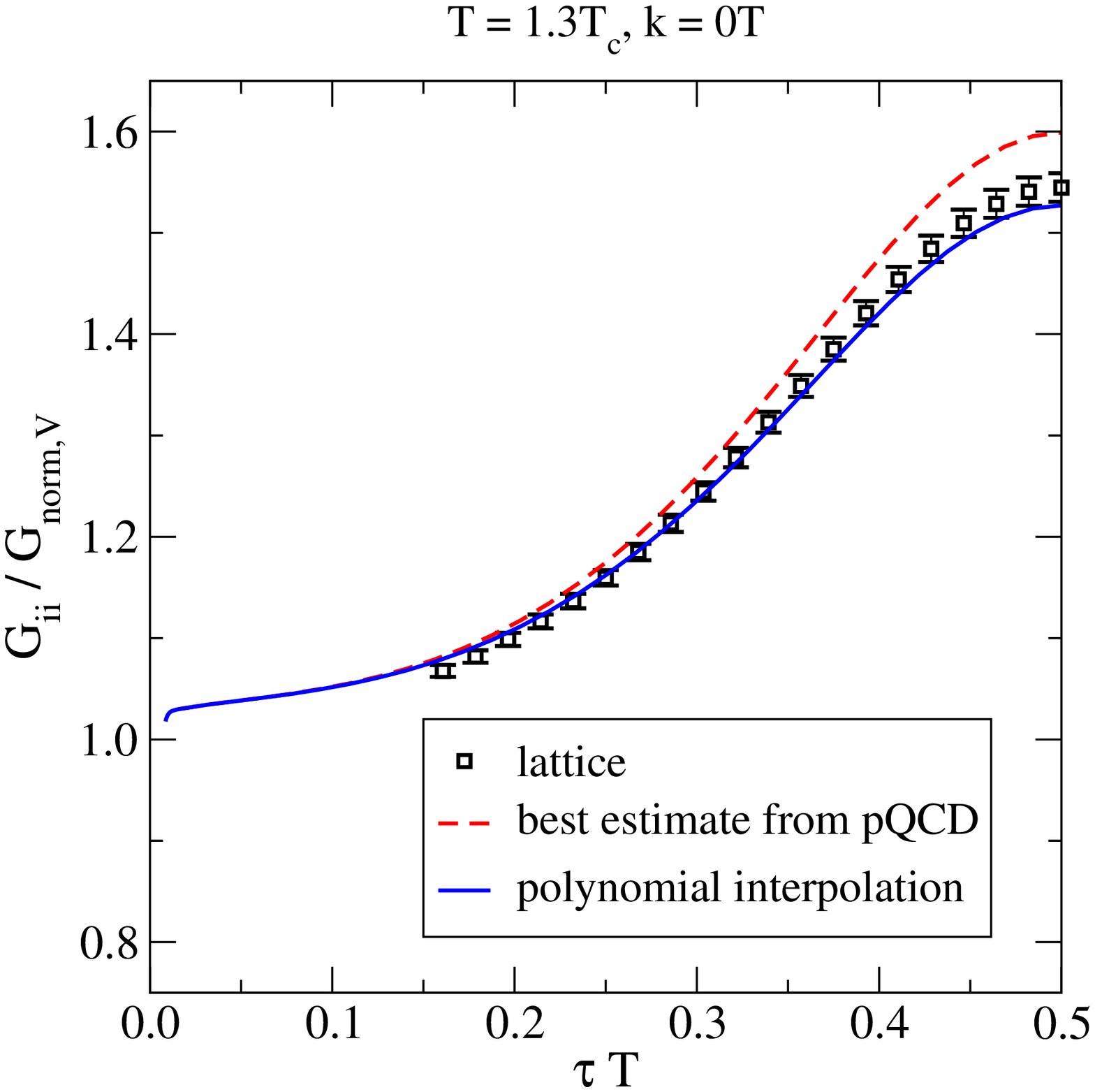}
% \hspace{0.1cm}
% \epsfysize=4.9cm\epsfbox{Gii_T14_k0.eps}
}

\caption[a]{\small
 Like in \fig\ref{fig:GV} but at zero momentum. 
 Only the spatial components of the vector 
 current have been included here. 
 The ``best estimate from pQCD'' %% (perturbative QCD)
 is based on refs.~\cite{nlo3,mr,cond4}, and has 
 been constructed as explained in footnote~\ref{fn:k0}.
}

\la{fig:Gii}
\end{figure}
%%%%%%%%%%%%%%%%%%%%%%%%%%%%%%%%%%%%%%%%%%%%%%%%%%%%%%%%%%%%%%%%%%%%%%%%%%%

%%%%%%%%%%%%%%%%%%%%%%%%%%%%%%%%% FIGURE %%%%%%%%%%%%%%%%%%%%%%%%%%%%%%%%%
\begin{figure}[t]

\hspace*{-0.1cm}
\centerline{%
 \epsfysize=7.5cm\epsfbox{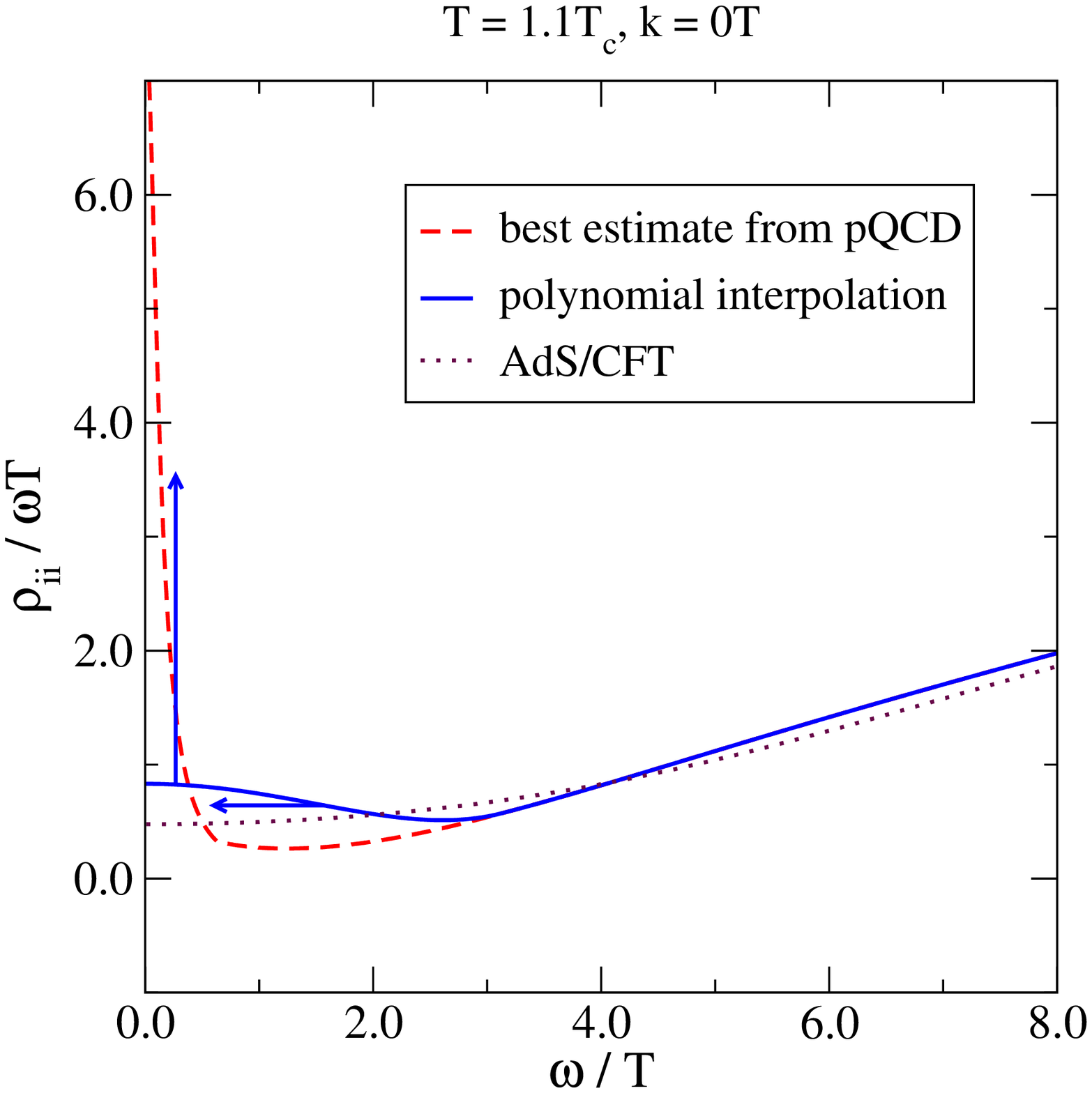}%
 \hspace{0.1cm}
 \epsfysize=7.5cm\epsfbox{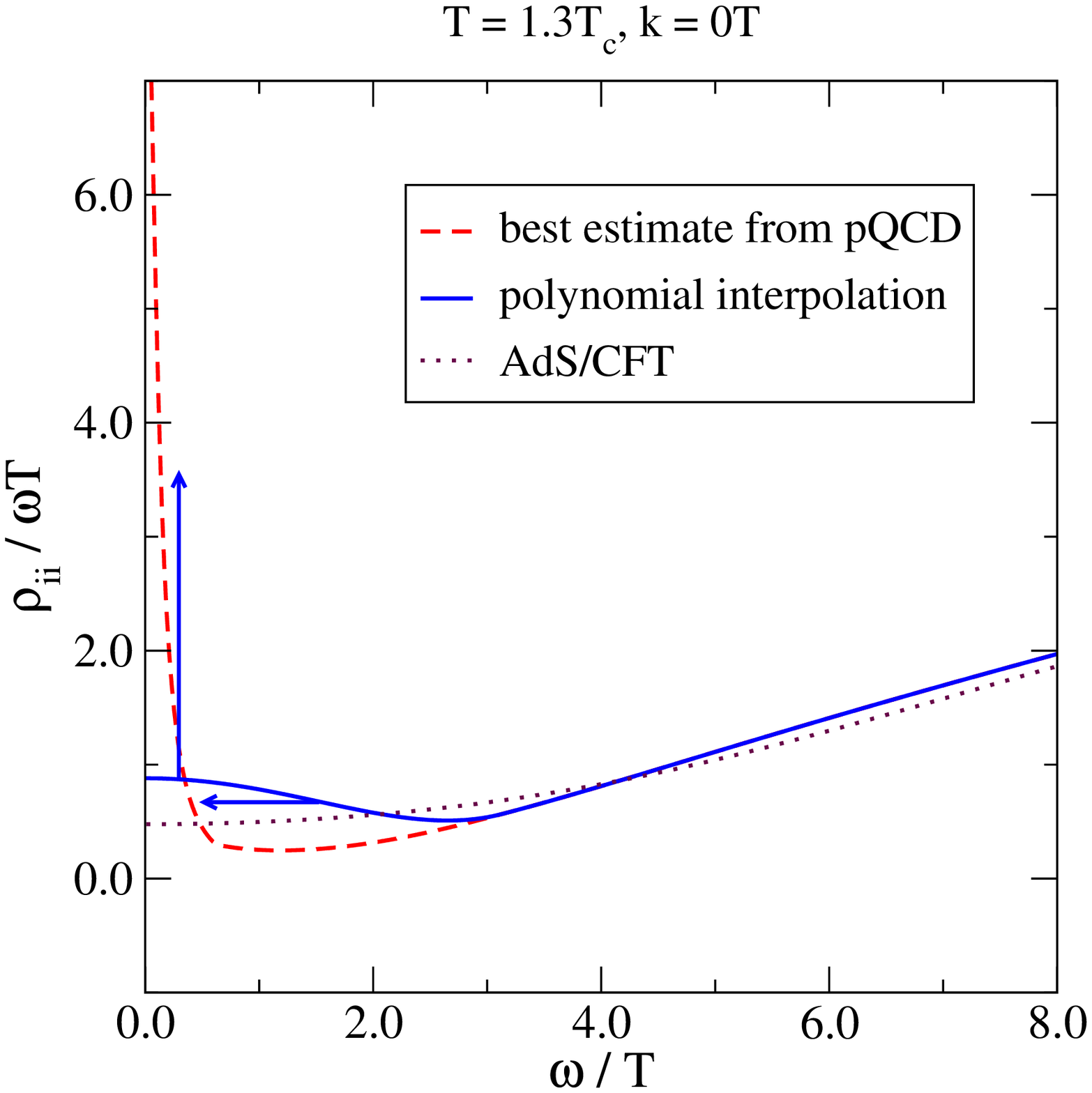}
% \hspace{0.1cm}
% \epsfysize=4.9cm\epsfbox{rho_T14_k0.eps}
}

\caption[a]{\small
 The spectral functions corresponding to \fig\ref{fig:Gii}
 from fits with $n_\rmi{max} = 0$. 
 The ``best estimate from pQCD''
 is based on refs.~\cite{nlo3,mr,cond4}, and has 
 been constructed as explained in footnote~\ref{fn:k0}.
% ref.~\cite{dilepton-lattice}.
 The AdS/CFT result %({\it to be added!}) 
 comes from ref.~\cite{ads2}, and has been rescaled
 to agree with the non-interacting QCD result at large $\omega/T$
 (cf.\ caption of \fig\ref{fig:rho_kn0}).
 As discussed in appendix~A and illustrated with the arrows, 
 the intercepts at $\omega = 0$ 
 are lower bounds, and the widths of the transport peaks
 are upper bounds. 
}

\la{fig:rho_k0}
\end{figure}
%%%%%%%%%%%%%%%%%%%%%%%%%%%%%%%%%%%%%%%%%%%%%%%%%%%%%%%%%%%%%%%%%%%%%%%%%%%

The extraction of transport coefficients at vanishing momentum, $k=0$, is 
faced with several challenges. 
One is that the transport peak could be
very narrow~\cite{mr} and therefore difficult to resolve from an
imaginary-time measurement. 
A separate problem is related to the domain of large frequencies, 
whose insufficient treatment may ``contaminate'' the extraction
of spectral features at low frequencies~\cite{cond4}. The latter
problem can be alleviated by making use of similar methods as discussed
in the main body of our paper. 
Numerical ``best estimate from pQCD'' spectral functions 
that can be used for this purpose, 
based on refs.~\cite{nlo3,mr,cond4},  
have been tabulated in ref.~\cite{dilepton-lattice}.\footnote{%
 More precisely, these results have been obtained by combining 
 the Born result with N$^4$LO vacuum corrections~\cite{kit_ns,kit_si}, 
 valid for $\omega \gg \pi T$~\cite{ope}, 
 with the NLO result valid for
 $\omega\sim \pi T$ \cite{nlo3}, and then taking
 the largest between this combination and the $\omega\sim \alphas^2 T$
 result~\cite{mr}, featuring a perturbative transport peak. 
 In this way at small $\omega$ we obtain
 a transport peak, at intermediate $\omega$ the LO+NLO sum, and 
 at large $\omega$ the N$^4$LO asymptotics.
 We have checked that the results of 
 ref.~\cite{nlo3} agree with the 
 $k\to0$ limit of the NLO correction in ref.~\cite{dilepton}, once the partial
 resummation of the thermal mass performed in ref.~\cite{nlo3} is undone, 
 being unjustified for $\omega \gsim \pi T$. 
 For what concerns
 the transport peak, we have ``quenched'' the calculation of
 ref.~\cite{mr} by removing $2\leftrightarrow 2$ processes with
 more than 2 external fermion lines from the collision operator
 and by fixing the Debye mass to its $\Nf=0$ value. \la{fn:k0}
 }
In this appendix we show the results that we obtain 
if the small-frequency domain is subsequently
modelled through \eq\nr{fit}. 

Like at non-zero momentum, the procedure described leads to 
a reasonably good description of the imaginary-time correlators 
at $\tau T \gsim 0.2$ ($\chi^2/$d.o.f.\ $\gsim 1.4$). 
This is illustrated in \fig\ref{fig:Gii}, with the corresponding
spectral functions shown in \fig\ref{fig:rho_k0}. The  
diffusion coefficient, defined through \eq\nr{Kubo_D}, is 
displayed in \fig\ref{fig:Deff}, as obtained from the
bootstrap sample with $n_\rmi{max} = 1$. 

It must be stressed, however, that our results at $k=0$ suffer from 
substantial systematic uncertainties. Indeed, 
if we fix $n_\rmi{max}=0$ and vary the fitting
point, like in \fig\ref{fig:chisq}, then $\chi^2$/d.o.f.\ does not show
a minimum but rather increases as a function of $\omega_0$. 
It is rather flat for $\omega_0 \lsim T$, however
then the transport peak is narrower than shown in 
\fig\ref{fig:rho_k0} and correspondingly the value of
the intercept at $\omega = 0$ is larger (the area under
the transport peak remains roughly constant). More quantitatively, 
values up to 
$\rho^{ }_{ii}/(\omega T) \lsim 4.5$ can be
obtained with $\chi^2$/d.o.f.\ $\sim 1.4$ for $\omega_0 \lsim T$; 
this corresponds to $DT \lsim 1.8$. We conclude that
the narrowness of the transport peak at $k=0$ poses a formidable
challenge which is not solved by our approach. 
Finally we remark that in a companion paper~\cite{setup} 
different ans\"atze led to the estimates 
$\rho^{ }_{ii}/(\omega T) \sim 0.6 ... 2.1$ at $T = 1.1\Tc$
and  $\rho^{ }_{ii}/(\omega T) \sim 0.6 ... 1.2$ at $T = 1.3\Tc$, 
which are quite consistent with \fig\ref{fig:rho_k0}
(note that the normalization of $\rho^{ }_{ii}$ in ref.~\cite{setup} 
differs by a factor 2 from the present paper).

%%%%%%%%%%%%%%%%%%%%%%%%%%%%%%%%%%%%%%%%%%%%%%%%%%%%%%%%%%%%%%%%%%%%%%%%%%%
%

\end{document}